\documentclass[aps,12pt,a4paper]{revtex4}
\usepackage{graphicx}
\usepackage{amsmath,amssymb}

\begin{document}

\title{Kinetic Energy Release in Fragmentation Processes following Electron Emission:
A time dependent approach}

\author{Ying-Chih Chiang$^1$, Frank Otto$^1$, Hans-Dieter Meyer$^1$, Lorenz S. Cederbaum$^1$}

\affiliation{$^1$ Theoretische Chemie, Universit\"at Heidelberg,
    Im Neuenheimer Feld 229, D--69120 Heidelberg, Germany}

\date{\today}

\begin{abstract}
A time-dependent approach for the kinetic energy release (KER) spectrum is developed
for a fragmentation of a diatomic molecule after an electronic decay process, e.g. Auger process. 
It allows one to simulate the time-resolved spectra and provides more insight into
the molecular dynamics than the time-independent approach.
Detailed analysis of the time-resolved emitted electron and KER spectra
sheds light on the interrelation between wave packet dynamics and spectra.  
\end{abstract}

\maketitle

\section{Introduction} 
\label{sec:intro}
Due to advances in time-of-flight mass spectroscopy \cite{Nasrin08,Jagutzki02}, 
measuring the velocity distribution of charged particles emerged into an important 
analysis tool during the last two decades.  
In consequence, fragmentation processes involving ions have been extensively studied, 
e.g. photodissociation \cite{Schinke} or Coulomb explosion \cite{Purnell94,Stapelfeldt95,Lezius98}.
Among different types of dissociation, one
interesting category is an electronic decay process (autoionization) followed by fragmentation, 
e.g. molecular Auger \cite{Eberhardt87,Weber03}
or interatomic coulombic decay (ICD) \cite{Lenz97,Robin00,Marburger03,Jahnke04,Nico10_1} processes. 
It is well known that removing a core electron
from a molecule leads to an Auger process and a doubly
ionized state is produced. This doubly ionized state 
is usually repulsive, but sometimes the molecular bond
is strong enough to overcome the Coulomb repulsion
and forms a temporarily bound final state \cite{Eberhardt87,Weber03},
i.e. a resonance.
On the other hand, removing an inner-valence electron
from a van der Waals molecule allows the ionic dimer to
further emit an outer-valence electron 
(ICD electron) and to produce two ions, which
then undergo Coulomb explosion since the
dimer is weakly bound \cite{Robin00,Jahnke04,Nico10_1}. 
In these two cases, one can measure the kinetic energy distribution of the
emitted Auger/ICD electron (electron spectrum) and the kinetic energy release
(KER) which is the sum of kinetic energies of all fragment ions 
\cite{Eberhardt87,Weber03,Weber01,Jahnke04}.
Due to the fact that not all fragments are charged in fragmentation
processes after resonance Auger processes \cite{Morin86,Menzel96,Elke98,Faris99},
we shall concentrate on the usual Auger processes.

For a diatomic system, the electron and KER spectra 
are usually considered to be mirror images of each other \cite{Jahnke04,Simona04}.
This ``mirror image principle" is based on energy conservation
arguments within a classical picture.
Suppose that the system has a total energy $E_T$ before emitting an
electron.  After emitting an electron with kinetic energy $E_e$, 
the system starts to dissociate, and at the end, all ionic fragments have a total
kinetic energy $E_{\text{KER}}$ and a final potential energy $V_f^{\infty}$
in the asymptotic region.
The energy conservation requires $E_T=E_e+E_{\text{KER}}+V_f^{\infty}$.
This relation gives a one-to-one mapping between
$E_e$ and $E_{\text{KER}}$ under the classical picture.
Hence the probability of measuring the electron with kinetic energy $E_e$
is equal to finding all fragments with the total kinetic energy $E_{\text{KER}}$.
However, the mirror image principle does not take into
account the situation where $E_T$ is not uniquely defined, i.e. $E_T$
has an energy distribution of finite width instead of a delta function.
We have shown previously that this initial energy distribution 
can break the mirror image principle \cite{Chiang11},
and details will be discussed via two model studies in Sec.~\ref{sec:MIP}.

The fully time-dependent approach \cite{Chiang11} allows to trace
how the spectrum evolves in time and to relate
this spectral development with the nuclear motion.
In Sec. \ref{sec:theory}, we will explain explicitly the theory for calculating 
the time-resolved KER spectrum with application to 
a fragmentation following electronic decay.  
Our theory of KER distribution is not limited 
to such cases but can easily be adapted to pump-probe experiments,
where the time-resolved KER spectrum has been measured 
to trace the molecular motion \cite{Jiang10_2,Bocharova11} .   
In Sec. \ref{sec:MIP}, we utilize two models to explain shortcomings 
of the mirror image principle.
As a highlight of the time-dependent approach, we compare in Sec. \ref{sec:TDKER} 
the time-resolved KER spectrum in detail with the time-resolved electron spectrum
using realistic potential curves.


\section{Theory} 
\label{sec:theory}

\begin{figure}[b]
\centering 
\includegraphics[width=8cm]{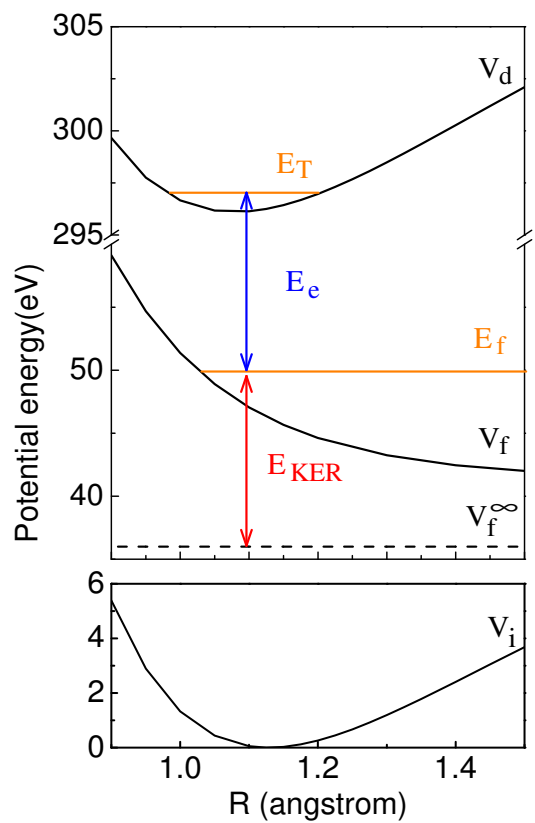} 
\caption{(Color online) Schematic energy diagram
for the molecular Auger process
$\text{CO}^{+} \to \text{CO}^{2+} (\;^3 \Sigma^-) +e^-$.
The internuclear distance is denoted by $R$. 
At $t=0$ the wave packet starts propagation on potential $V_d$
with a total energy $E_T$, 
and it gradually decays to the repulsive potential $V_f$
via emitting an electron with kinetic energy $E_e$ and
producing the kinetic energy release $E_{\text{KER}}$. 
Potential curves are taken from Ref.~\cite{Eland04,Carroll02} .
}
\label{fig:COPEC}
\end{figure}

A schematic picture of fragmentation following autoionization is
depicted in Fig.~\ref{fig:COPEC}. Before the pulse 
arrives, a molecule is in its electronic ground state ``$i$''.
After ionizing a photoelectron from the molecule (or photoexciting the molecule), the
intermediate state ``$d$'' is populated. Then the system 
decays to the repulsive final state ``$f$'' via electron emission.
The potential energy curve of state $i$ is denoted as $V_i$, and so on.
The emitted electron (Auger or ICD electron) has a 
kinetic energy $E_e$ and the fragmented ions have
a total kinetic energy $E_{\text{KER}}$ after the dissociation
completes. 

The ionic intermediate state $d$, accessed via photoionization, 
is chosen as an example, see Ref. \cite{Elke96} for an excitation scenario.
To describe such a process, the equations of motion can be obtained 
through a procedure described in Ref. \cite{Elke96}. 
The total wave function ansatz for multiple final states reads
\begin{equation}
\label{eq:psitot}
|\Psi_\text{tot}(t)\rangle =
|\psi_i(t)\rangle |\phi_i\rangle \,+\,
|\psi_d(E_{\text{ph}},t)\rangle |\phi_d , E_{\text{ph}}\rangle \,+\,
\sum_f \int \!\!\! \int \text{d} E_{\text{ph}} \; \text{d}E_e \; |\psi_f(E_{\text{ph}},E_e,t)\rangle
|\phi_f,E_{\text{ph}},E_e\rangle
\:,
\end{equation}
where $|\phi_i\rangle$ and $|\psi_i\rangle$ denote the 
electronic and nuclear wave function of state $i$,
respectively, and likewise for states $d$ and $f$.
The symbol $|\phi_f,E_{\text{ph}},E_e\rangle$ denotes the electronic
wave function of the di-cationic state $f$ augmented with
the wave function of the photoelectron $E_{\text{ph}}$ and the emitted Auger or ICD electron of energy $E_e$.
Inserting the total wave function and the total Hamiltonian \cite{Elke96}
into the time-dependent Schr\"odinger equation and projecting onto
the electronic wave functions, our effective working equations read 
\begin{align}
i |\dot{\psi}_{i}(t)\rangle &= \hat{H}_i |\psi_{i}(t)\rangle + \int \text{d} E_{\text{ph}} \hat{F}^\dagger(E_{\text{ph}},t) |\psi_d(E_{\text{ph}},t)\rangle 
\label{eq:ioneqi}
\\
i |\dot{\psi}_{d}(E_{\text{ph}},t)\rangle &= \hat{F}(E_{\text{ph}},t) |\psi_i(t)\rangle + (\hat{H}_d + E_{\text{ph}} - \tfrac{i}{2} \Gamma) |\psi_{d}(E_{\text{ph}},t)\rangle
\label{eq:ioneqd}
\\
i |\dot{\psi}_f(E_{\text{ph}},E_e,t)\rangle &= W_{d \to f}|\psi_{d}(E_{\text{ph}},t)\rangle + (\hat{H}_f + E_{\text{ph}}+E_e)|\psi_f(E_{\text{ph}},E_e,t)\rangle
\label{eq:ioneqf}
\end{align}
where $\hat{H}_i$ is the nuclear Hamiltonian for state $i$.
The transition matrix element $W_{d \to f}$ 
is closely related to the total decay rate $\Gamma$ based on the
local approximation by $\Gamma=2\pi \sum_f |W_{d \to f}|^2$ \cite{Elke96}. 
The interaction with the external pulse is given by
$\hat{F}(E_{\text{ph}},t)= \langle \phi_d, E_{\text{ph}}|\hat{\mathbf{D}}|\phi_i \rangle \cdot \mathbf{E} (t)$,
where $\mathbf{E} (t)$ denotes the external electric field. 
After photoionization, the photoelectron can 
carry an arbitrary amount of kinetic energy
due to the fact that most experiments utilize photons with
energy well above the threshold.
Even photoionization caused by a coherent light source, e.g. synchrotron radiation, 
will therefore populate different vibrational levels
of state $d$, according to the Franck-Condon principle, see e.g. Ref. \cite{Matsumoto06}.
In other words, the outgoing photoelectron and remaining intermediate state $d$
together obey energy conservation, but
the energy of both these subsystems will be subject to a certain
energy distribution.  The initial condition of an ionization process, 
hence, is equivalent to a broad-band excitation, which adopts
$F(t)\sim \delta (t)$ and thus places an initial wave packet vertically on $V_d$.
This initial condition allows us to concentrate only on the dynamics
of state $d$ and $f$, and $E_{\text{ph}}$ is merely shifting the
potential energy for both states without changing the norm
of the wave packet.  
Since we are interested only in the ionic molecule which
will undergo autoionization and fragmentation, the  
dummy $E_{\text{ph}}$ can be removed and the effective working equations
read \cite{Elke96} :
\begin{align}
i |\dot{\psi}_{d}(t)\rangle &= \hat{F}(t) |\psi_i(t)\rangle + (\hat{H}_d - \tfrac{i}{2} \Gamma) |\psi_{d}(t)\rangle
\label{eq:mastereqd}
\\
i |\dot{\psi}_f(E_e,t)\rangle &= W_{d \to f}|\psi_{d}(t)\rangle + (\hat{H}_f + E_e)|\psi_f(E_e,t)\rangle
\label{eq:mastereqf}
\end{align}

A bridge between the time-dependent and time-independent
approach is built by expanding the wave packet in sets of eigenfunctions.  
For example, the nuclear components $|\psi_d(t)\rangle$ and
$|\psi_f(E_e,t)\rangle$ can be expanded
as follows:
\begin{align}
|\psi_d(t)\rangle = \sum_{n_d} c_{n_d}(t) |n_d\rangle
\quad;\quad
|\psi_f(E_e,t)\rangle = \int\!\text{d}E_f\:c_{E_f}(E_e,t) |E_f\rangle
\label{eq:exp-psi}
\end{align}
where $|n_d\rangle$ and $|E_f\rangle$ are the eigenfunction of nuclear Hamiltonians,
\begin{align}
(\hat{H}_d - \tfrac{i}{2} \hat{\Gamma}_d) |n_d\rangle &= \varepsilon_{n_d} |n_d \rangle
\quad;\quad
\varepsilon_{n_d} = E_{n_d} - \tfrac{i}{2} \Gamma_{n_d}
\label{eq:defnd}
\\
\hat{H}_f |E_f\rangle &= E_f |E_f\rangle
\label{eq:defnf}
\:.
\end{align}
The Hamiltonian of the intermediate state $d$ is a non-Hermitian, complex-symmetric one.
Its eigenfunctions are orthonormal with respect to the symmetric scalar product
$( \cdot | \cdot )$, i.e. $\langle n_d^*|m_d\rangle = (n_d| m_d)=\delta_{n_d,m_d}$.
Note that $(f|g)=\int f(x)g(x) \, \text{d}x$ is originally a mathematical notation 
for the symmetric scalar product, but here it motivates us to define the state notation $|n_d)=|n_d\rangle$
and $(n_d|=\langle n_d^*|$, which extends the usual Dirac notation.
We emphasize that the final state potential $V_f$ is purely repulsive. Its eigenfunctions form
a continuum basis set with normalization
\begin{equation}
\langle E_f|E_f' \rangle = \delta(E_f-E_f') \:.
\label{eq:Ef_norm}
\end{equation}
For processes depicted in Fig.~\ref{fig:COPEC},
one can measure the electron spectra and the KER spectra,
either separately or in coincidence.
There are also two different numerical
methods to obtain the spectra.
One is through evaluating the coincidence spectrum:
integrating the coincidence spectrum over $E_e$ gives the KER spectrum,
and vice versa.
Let us start from the coincidence spectrum.
It measures the probability to find an electron with kinetic energy $E_e$ 
and the fragments with a total kinetic energy $E_{\text{KER}}$ 
in coincidence.
These conditions define a projector $\hat{P}$, which when evaluated with the
total wave function $\Psi_\text{tot}$ yields the coincidence probability.
As we know that $E_\text{KER}$ is related to the final state 
nuclear energy $E_f$ by $E_\text{KER}=E_f - V_f^{\infty}$,
this projector reads
\begin{equation}
\label{eq:projector}
\hat{P}=|E_f \rangle \langle E_f| \otimes |\phi_f \;, E_e \rangle \langle \phi_f \;, E_e | \:.
\end{equation}
The time-resolved coincidence spectrum, of course, 
is obtained straightforwardly.
\begin{equation}
\label{eq:tdcoincidence}
\sigma(E_{\text{KER}},E_e,t)=\langle \Psi_{\text{tot}} | \hat{P} | \Psi_{\text{tot}} \rangle
= |\,\langle E_f | \psi_f(E_e,t) \rangle\,|^2 = |c_{E_f}(E_e,t)|^2
\end{equation}
After evaluating the time-resolved coincidence spectrum, the 
time-resolved KER and electron spectra are also available
from integrating Eq.~(\ref{eq:tdcoincidence}) numerically \cite{Nico10_2}:
\begin{equation}
\label{eq:tdkerviaco}
\sigma_{\text{KER}} (E_{\text{KER}},t) = \int_0^\infty\!\text{d}E_e\;\sigma(E_{KER},E_e,t)
\end{equation}
\begin{equation}
\label{eq:tdespectviaco}
\sigma_{e}(E_e,t) = \int_0^\infty\!\text{d}E_{\text{KER}}\;\sigma(E_{KER},E_e,t) .
\end{equation}
As one can see, Eqs.~(\ref{eq:tdkerviaco},\ref{eq:tdespectviaco}) require a numerical
integration over the coincidence spectrum.  The idea is simple, but this way often takes 
enormous numerical effort to obtain a KER or electron spectrum for comparing with experiments,
because it requires a very fine energy mesh to do these integrals accurately.
Besides, the physical meaning of the KER and electron spectra 
is obscured under this representation.

A better way is to carry out these integrations analytically,
avoiding evaluation of the coincidence spectrum.
To simplify the mathematics, we temporarily set the zero
point of the energy such that $V_f^{\infty}=0$ and hence $E_{\text{KER}}=E_f$.
Eq.~(\ref{eq:tdespectviaco}) can then be rearranged to a different form
by inserting another integral and using the orthonormal property from Eq.~(\ref{eq:Ef_norm}):
\begin{align}
\label{eq:tdespect}
\sigma_{e}(E_e,t) &= \int_0^\infty\!\text{d}E_f \; |c_{E_f}(E_e,t)|^2 
\nonumber \\
&= \int_0^\infty\!\text{d}E_f \;  \int_0^\infty\!\text{d}E'_{f}\; c^*_{E'_{f}}(E_e,t)  \langle E'_{f} | E_f \rangle c_{E_{f}}(E_e,t)
\nonumber \\
&= \langle \psi_f (E_e,t) |\psi_f (E_e,t) \rangle
\end{align}
This gives the well-known result of the time-resolved  electron spectrum \cite{Elke96}
as expected.  It takes, however, more effort to obtain the equation for
the time-resolved KER spectrum.  We assume that the potential curves $V_d$ and $V_f$
are well separated, so that the coincidence probability $|c_{E_f}(E_e,t)|^2$
is zero when $E_e$ becomes negative. This allows the lower bound of Eq.~(\ref{eq:tdkerviaco})
to be extended to $- \infty$. 
In addition, from Eqs.~(\ref{eq:mastereqf}, \ref{eq:exp-psi}) follows
\begin{equation}
\label{eq:coefficient_f}
c_{E_f} (E_e,t) = -i \int_0^t e^{i(E_f+E_e)(t'-t)} \langle E_f | \hat{W}_{d \to f} | \psi_d (t') \rangle \, \text{d} t'   \; .
\end{equation}
Inserting this expression into Eq.~(\ref{eq:tdkerviaco}),
we arrive at another form of the time-resolved KER distribution:
\begin{align}
\label{eq:tdker}
&\sigma_{\text{KER}}  (E_{\text{KER}},t) 
\nonumber \\
&= \int_0^t\!\text{d}t' \int_0^t\!\text{d}t'' \int_{-\infty}^{\infty}\!\text{d}E_e \;
e^{i(E_f+E_e)(t''-t')} \; \langle \psi_d (t') | \hat{W}^+_{d \to f} | E_f\rangle
   \langle E_f | \hat{W}_{d \to f} | \psi_d (t'') \rangle
 \nonumber \\
&= \int_0^t\!\text{d}t' \int_0^t\!\text{d}t'' \; 2 \pi \delta(t''-t') 
\langle \psi_d (t') | \hat{W}^+_{d \to f} | E_f \rangle \langle E_f| \hat{W}_{d \to f} | \psi_d(t'') \rangle 
\nonumber \\
&= 2 \pi \int_0^t \!\text{d}t' \; | \langle E_f |\hat{W}_{d \to f}| \psi_d (t') \rangle |^2\; . 
\end{align}
This new formulation provides a vivid physical interpretation
of the KER spectrum \cite{Chiang11} as an ``accumulated" generalized Franck-Condon factor.
(``Generalized'' because the usual Franck-Condon factor assumes that
$\hat{W}_{d \to f}$ is independent of $R$.)
This interpretation differs remarkably from the one of the electron
spectrum, cf. Eq.~(\ref{eq:tdespect}), as the latter depends only on $|\psi_f\rangle$.
Note that Eq.~(\ref{eq:tdker}) is also valid for state $d$ which is accessed by excitation.

The KER spectrum can also be interpreted as the 
accumulated amount of the wave packet  $|\psi_d(t)\rangle$
mapped onto an eigenfunction of a final electronic state
by the following operator
\begin{equation}
\label{eq:keroperator}
\hat{O}_{\text{KER}} = 2 \pi \hat{W}^+_{d \to f} |E_f\rangle  \langle E_f| \hat{W}_{d \to f} \; . 
\end{equation}
Eq.~(\ref{eq:keroperator}) indicates that the mapping is weighted 
by the transition matrix element between two electronic states. 
Before entering the next section, we emphasize that 
Eqs.~(\ref{eq:tdcoincidence},\ref{eq:tdespect},\ref{eq:tdker}) are all
time-resolved spectra.  The static result (time-independent spectrum) is given by 
the limit $t \to \infty$.  When letting $t \to \infty$, our equations are hence equivalent to 
those derived from scattering theory \cite{Beck00,Simona03}.  
More details are given in the appendix.
In appendix B, we discuss how the KER spectrum, Eq.~(\ref{eq:tdker}), is evaluated
efficiently in a numerical stable manner.

 

\section{Breakdown of Mirror Image Principle} 
\label{sec:MIP}

Although the electron and KER spectra, see Eqs.~(\ref{eq:tdespect},\ref{eq:tdker}), 
are different, one may still wonder how the initial energy distribution
of intermediate state $d$ plays a role in the two, physically distinct, spectra?
To understand why the mirror image principle is violated when
there is an energy distribution of state $d$,
we will discuss two model situations with a single
final state $f$ and a constant $\Gamma$ (and thus constant $W_{d \to f}$).   
The first one, which we will call ``one-level model'', assumes that only
one vibrational eigenfunction $|n_d\rangle$ of the state $d$ is 
initially populated.  Instead of traveling back and forth
on $V_d$, in this case $|\psi_d (t) \rangle$ remains at 
the same position and decays.  
For this model, the coincidence spectrum is 
easily derived from Eqs.~(\ref{eq:coefficient_f}) and (\ref{eq:tdcoincidence}) 
under the limit $t \to \infty$ 
\begin{equation}
\sigma(E_\text{KER},E_e) = \frac{ |\langle E_f | n_{d} \rangle|^2 \, \Gamma/2\pi}%
{(E_\text{KER}+V_f^\infty+E_e-E_{n_{d}})^2 + \Gamma^2/4}
\:.
\end{equation}
Integrating over the coincidence spectrum, the KER spectrum then results in (using
$E':=E_\text{KER}+V_f^\infty+E_e-E_{n_{d}})$
\begin{align}
\sigma_\text{KER}(E_\text{KER}) &= \int^\infty_{E_\text{KER}-(E_{n_{d}}-V_f^\infty)}
\text{d}E'\:\frac{\Gamma/2\pi}{E'^2+\Gamma^2/4}\,|\langle E_f | n_{d}\rangle|^2
\nonumber  \\ 
&= |\langle E_f | n_{d}\rangle|^2 \left( \frac{1}{2} +
\frac{1}{\pi} \arctan \frac{E_{n_{d}}-V_f^\infty-E_\text{KER}}{\Gamma/2} \right)
\:.
\end{align}
From Fig.~\ref{fig:COPEC} one can see that $E_{n_{d}}-V_f^\infty-E_\text{KER}$ 
is usually much larger than $\Gamma$, which allows to approximate
the $\arctan$ by its limit $\tfrac{\pi}{2}$ at $+\infty$:
\begin{equation}
\sigma_\text{KER}(E_\text{KER}) = |\langle E_f | n_{d}\rangle|^2
\end{equation}
Thus, in this case the KER spectrum  is 
simply given by the Franck-Condon factor. 
If we would alter only $\Gamma$ as a parameter,
the KER distribution would remain the same since the 
Frank-Condon factor is not influenced by the decay rate.

On the other hand,  Eq.~(\ref{eq:tdespectviaco})
under the limit $t \to \infty$ now leads to
\begin{align}
\sigma_e(E_e) 
&= \int^\infty_{E_e-(E_{n_{d}}-V_f^\infty)}
\text{d}E'\:\frac{\Gamma/2\pi}{E'^2+\Gamma^2/4}\,|\langle E_f | n_{d}\rangle|^2
\nonumber \\
&\approx \int_{-\infty}^{\infty}\!\!\text{d}E'\:
\frac{\Gamma/2\pi}{E'^2+\Gamma^2/4}\,
\sigma_\text{KER}(E'+E_{n_{d}}-V_f^\infty-E_e)
\nonumber\\
&= \int_{-\infty}^{\infty}\!\!\text{d}E'\:
\frac{\Gamma/2\pi}{E'^2+\Gamma^2/4}\,
\sigma_\text{mKER}(E_e-E')
\:.
\label{eq:spec-el1}
\end{align}
where $\sigma_\text{mKER}(E_e)$ denotes the mirror image of the
KER spectrum,
\begin{equation}
\sigma_\text{mKER}(E_e) = \sigma_\text{KER}((E_{n_{d}}-V_f^\infty)-E_e)
\label{eq:mker}
\end{equation}
Eq.~(\ref{eq:spec-el1}) shows that the electron spectrum is 
obtained by convoluting the mirrored KER spectrum with a Lorentzian 
of width $\Gamma$ (FWHM).
In other words, the finite lifetime of the intermediate state $d$ 
broadens the electron spectrum but not the KER spectrum.
When the Lorentzian has a width approaching zero, 
it becomes a delta function and the convolution of Eq.~(\ref{eq:spec-el1})
states that the mirror image relation between 
the electron and KER spectra holds precisely.

\begin{figure}
\centering 
\includegraphics[width=8cm,height=6cm]{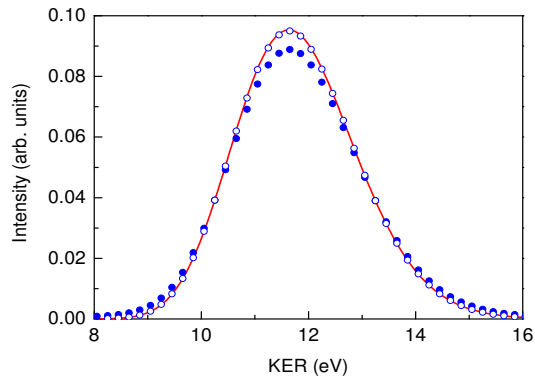} 
\vspace{-0.5cm}

\caption{(Color online) KER spectrum and
mirror image of electron spectra for the one-level model.
The initial wave packet is chosen to be the ground vibrational
eigenfunction of state $d$.  
Potential curves are shown in Fig.~\ref{fig:COPEC}. 
$\Gamma$ is chosen to be either
9.5 or 200 meV.  The numerical KER spectrum (red curve) 
is always the same regardless of $\Gamma$.
However, the mirror image of the electron spectrum
only coincides with the KER spectrum when
$\Gamma$ is small (open circles).
For $\Gamma=200$ meV, the mirror
image of the electron spectrum (blue circles)
is broader than the KER spectrum.
}
\label{fig:plotCOvib0}
\end{figure}
The numerical result for the KER spectrum is plotted with the red line
in Fig.~\ref{fig:plotCOvib0}.  Potential curves, shown in Fig.~\ref{fig:COPEC}, 
are taken from a real example \cite{Eland04,Carroll02}: 
the intermediate state $\text{CO}^+ (\text{C} \; \text{1s}^{-1})$
and the repulsive final state $\;^3\Sigma^{-}$.
The real Auger width for this intermediate state is 95 meV \cite{Carroll02},
but we take 200 meV (roughly twice) and 9.5 meV (one tenth)
in order to show the physics clearly.
For a small $\Gamma$ like 9.5 meV, the mirror
image of the electron spectrum coincides with
the KER spectrum, see the open circles and 
the red curve.  In contrast, the mirror image 
principle breaks down in the case of
a large $\Gamma$, see the blue circles and
the red curve, due to the finite lifetime broadening.  
Thus, the first restriction for the validity
of the mirror image principle is that the decay width 
$\Gamma$ has to be small.



\begin{figure}
\centering 
\includegraphics[width=8cm,height=6cm]{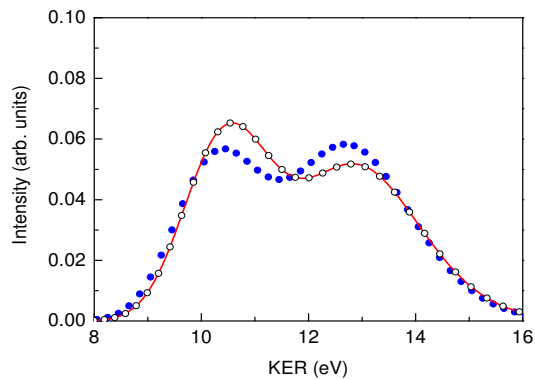} 
\vspace{-0.5cm}

\caption{(Color online) KER spectrum and
mirror image of electron spectra for the two-level model.
The initial wave packet is $\frac{1}{\sqrt{2}}(|n_0)+|n_1))$.
Potentials are the same as depicted in Fig.~\ref{fig:COPEC}. 
$\Gamma$ is chosen to be 9.5 meV, as it needs to be small
for the mirror image principle to hold at all.
Due to the very small $\Gamma$, the interference
between the two levels is so small that the KER
spectrum (red curve) is just a weighted sum of two KER spectra
whose initial wave packet are chosen to be either $|n_0)$ or $|n_1)$,
see the open circles.
The mirror image of the numerical electron spectrum
is plotted with blue circles and is clearly different from
the KER spectrum, even though $\Gamma$ is very small.}
\label{fig:plotCOvib01}
\end{figure}

Does a very small $\Gamma$ always guarantee the
validity of the mirror image principle? 
We continue our investigation with $\Gamma=9.5 \text{meV}$ and
the same potential curves. Only the initial
wave packet is now chosen to be a linear combination 
of two eigenfunctions of state $d$: $c_a|n_a)+c_b|n_b)$.
We will call this situation a ``two-level model''.  
As we already know from our study so far, the 
mirror image principle holds if only one of the 
eigenfunctions is populated initially, i.e.
when $c_a=0$ or $c_b=0$. One might expect
the mirror image principle to still hold
in a two-level model.  As an example, we define the 
initial wave packet to be a linear combination
of the first two eigenfunctions of state $d$, i.e.
$\frac{1}{\sqrt{2}}(|n_0)+|n_1))$, and the
numerical results of this two-level model
are shown in Fig.~\ref{fig:plotCOvib01}.     
The simulated KER spectrum (red curve) coincides
with the summation over individual KER spectra (open circles).
The interference between the two vibrational levels is negligible
because the $\Gamma$ is much smaller than the energy gap
between  the vibrational levels.  Although the interference is
small and the broadening of the lifetime is not obvious,
the mirror image of the electron spectrum (blue circles)
still differs from the KER spectrum.  This is due to the
``out-of-focus symmetry planes'' in this case. 
Recall that the energies are relates by $E_T=E_e+E_{\text{KER}}+V^{\infty}_f$,
which can be rearranged as $E_{\text{KER}}=(E_T-V^{\infty}_f)-E_e$.
For a one-level model, the symmetry plane is located
at the energy $(E_{n_d}-V^{\infty}_f)/2$.  However,
if two vibrational levels are populated initially,
the symmetry plane between KER and electron
spectra is located at a different energy
for each level, i.e.
at $(E_a-V^{\infty}_f)/2$ for $|n_a)$ and  at $(E_b-V^{\infty}_f)/2$ for $|n_b)$.
If $\Gamma$ is small, the KER (electron) spectrum 
in the two-level model is the same as the weighted sum
of the individual KER (electron) spectra obtained via a one-level model each.
However, the weighted summation of the two spectra
actually destroys the symmetry plane, i.e. no symmetry plane
is left.  Therefore, the KER and electron spectra are different
in the two-level model.

The physics will be more transparent by showing the
analytical solutions of the two-level model.
Assume the initial wave packet is $c_a|n_a)+c_b|n_b)$; 
the corresponding coincidence spectrum reads
\begin{align}
\sigma(E_\text{KER},E_e)
 = \frac{\Gamma}{2\pi} \biggl(
&  \frac{|\langle E_f | n_a) |^2\,|c_a(0)|^2}{(E_f+E_e-E_a)^2 + \Gamma^2/4}
 + \frac{|\langle E_f | n_b) |^2\,|c_b(0)|^2}{(E_f+E_e-E_b)^2 + \Gamma^2/4}
\nonumber\\ &
 + 2\,\text{Re}\,\frac{c_b(0)^*\,c_a(0)\;(n_b|E_f\rangle \langle E_f|n_a)}
      {(E_f+E_e-E_a+\tfrac{i}{2}\Gamma)(E_f+E_e-E_b-\tfrac{i}{2}\Gamma)}
   \biggr)
\label{eq:spec-coin-twovib}
\end{align}
Using $\bar{E} = (E_a\!+\!E_b)/2$
and $\delta E= E_b\!-\!E_a$,
the interference term can be written as
\begin{equation}
2\,\text{Re}\,\frac{c_b(0)^*\,c_a(0)\;(n_b|E_f\rangle \langle E_f|n_a)}%
      {(E_f+E_e-\bar{E})^2 - (\delta E + i\Gamma)^2/4}
\end{equation}
Similar to the previous one-level model, 
integrating over the electron energy $E_e$ yields the KER spectrum:
\begin{align}
\sigma_\text{KER}(E_\text{KER})
&= |\langle E_f | n_a) |^2\,|c_a(0)|^2
\,+\, |\langle E_f | n_b) |^2\,|c_b(0)|^2
\nonumber\\
&+\,2\,\text{Re} \left[ c_b(0)^*\,c_a(0)\;(n_b|E_f\rangle \langle E_f|n_a)
\,\frac{1}{ 1 - i\delta E /\Gamma } \right]
\end{align}
Thus the KER spectrum is composed of two parts:  
weighted summation of the Franck-Condon
factors resulting from all vibrational levels, i.e. $|\langle E_f|n_a)|^2$
and $|\langle E_f|n_b)|^2$, plus the interference between different levels.
The latter contains cross-multiplied Franck-Condon overlaps and
a coefficient depending on $\Gamma$.
Why does the finite lifetime of state $d$ ,i.e. $1/\Gamma$, suddenly matter
in the KER spectrum?  Let us explain.  The coefficient's magnitude determines
how important the interference is.  In the two-level model,
$|\psi_d(t)\rangle$ decays over time, but it also travels back-and-forth 
between two turning points, with a period $2\pi/\delta E$.  
The ratio $\delta E/\Gamma$ is proportional to how many times 
the wave packet oscillates within its lifetime.
The larger this ratio, the smaller the contribution of the interference is.

For further reference, we will now denote the individual KER spectra 
caused by the different vibrational levels,
and their corresponding mirror image spectra, as
\begin{align}
\sigma_\text{KER}^{n_d}(E_\text{KER}) &:= |\langle E_f|n_{d})|^2\,\Bigl.\Bigr|_{E_f=E_\text{KER}+V_f^\infty}
\\
\sigma_\text{mKER}^{n_d}(E_e) &:= \sigma_\text{KER}^{n_d}((E_{n_d}-V_f^\infty)-E_e)
\label{eq:def-mker-nd}
\end{align}
As for the electron spectrum $\sigma_e$, the integration of Eq.~(\ref{eq:spec-coin-twovib})
over $E_\text{KER}$ is not easily possible, as the $|E_f\rangle$ depend on $E_\text{KER}$.
For very small $\Gamma$, the interference term can be neglected, and the integral can be
approximated in the same way as in the previous section, resulting in
\begin{align}
\sigma_e(E_e) &\approx
|c_{a}(0)|^2\,\Bigl. |\langle E_f | n_{a})|^2 \Bigr|_{E_f=E_a-E_e}+
|c_{b}(0)|^2\,\Bigl. |\langle E_f | n_{b})|^2 \Bigr|_{E_f=E_b-E_e}
\nonumber\\
&= |c_{a}(0)|^2\,\sigma_\text{mKER}^{n_{a}}(E_e)\,+\,
   |c_{b}(0)|^2\,\sigma_\text{mKER}^{n_{b}}(E_e) 
\:.
\end{align}
That is, the electron spectrum is given by the weighted sum of the mirrored KER spectra
of the individual vibrational levels. In general, this is \emph{not} the same as the
mirror image of the full KER spectrum, because the ``plane of mirror'' is different for
different vibrational levels (see Eq.~(\ref{eq:def-mker-nd})); again, this plane is at
$(E_{a}-V_f^\infty)/2$ for $|n_{a})$ and at $(E_{b}-V_f^\infty)/2$ for $|n_{b})$.
Only if the two vibrational energies are very close together, i.e. $E_{a}
\approx E_{b}$, the electron spectrum will be given in good approximation by
$\sigma_e(E_e) \approx \sigma_\text{KER}((\bar{E}-V_f^\infty)-E_e)$.
Hence the second requirement for the validity of the mirror image principle
is that the populated levels are quasi-degenerate.

From these two model examples,  we obtain the conclusion 
that the initial energy distribution of the state $d$ has to be close to a delta function,
i.e. its energy has to be defined rather sharply.
This is usually not true for a molecular Auger process, but the mirror 
image relation often holds for ICD in noble gas dimers
due to the weak Van der Waals interaction, see also \cite{Chiang11}.

\section{Time-Resolved KER and Electron Spectra} 
\label{sec:TDKER}

KER and electron spectra are very different: The KER spectrum contains the
information on the nuclear dynamics of the intermediate state $d$ while the
electron spectrum shows the population of the final state $f$.
Yet both spectra can be used for tracing the molecular motion
through measuring the time-resolved spectra.  
In this section, the differences between these two 
spectra will be illustrated by showing their evolution in time.
We choose a realistic case of $\text{C}^+\text{O}^+$
fragmentation following the Auger decay of $\text{C}^+ (\text{1s}^{-1})$,
see Fig.~\ref{fig:COPEC}. 
As we already mentioned, the total Auger width is 95 meV 
for this process \cite{Carroll02}. 
Although this channel has a lower branching ratio, 
its KER distribution has been observed in experiments and
it was found to be responsible for the broad KER distribution above
10 eV \cite{Weber01}. 
%

\begin{figure}
\centering 
\includegraphics[width=8cm,height=5.5cm]{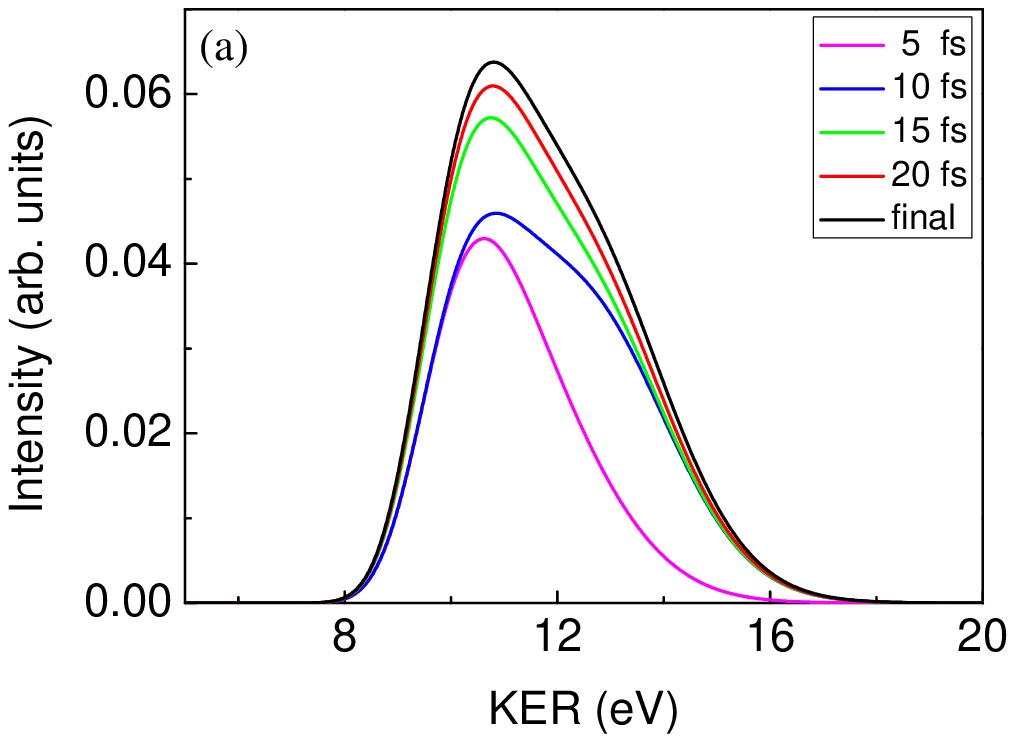} 
\vspace{-0.5cm}
\includegraphics[width=8cm,height=5.5cm]{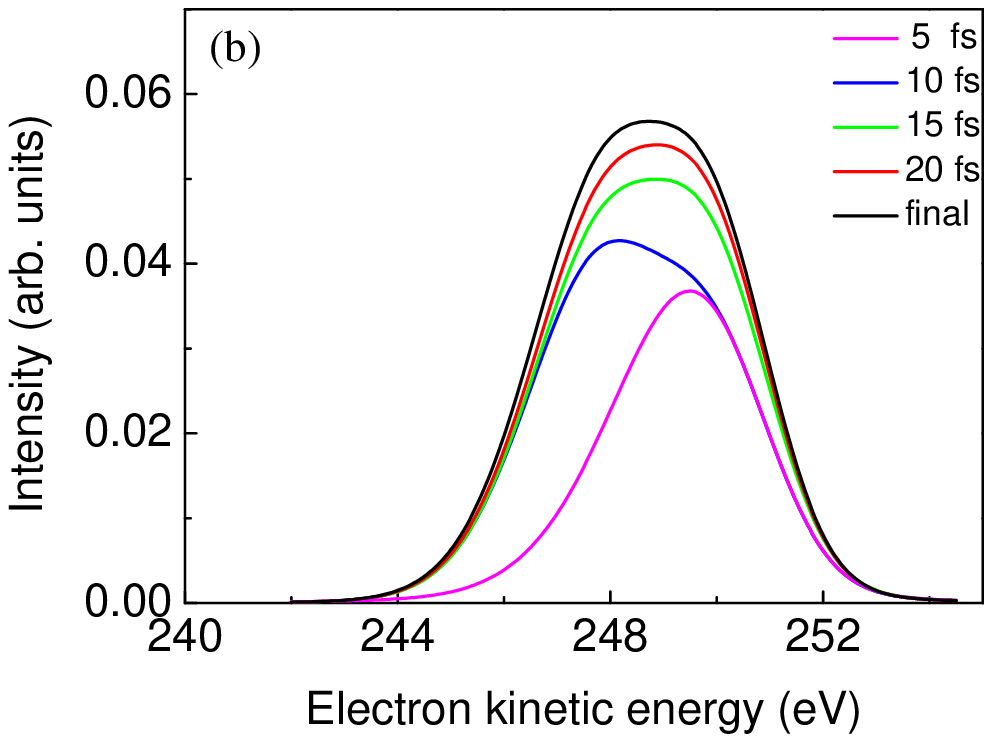} 
\vspace{-0.5cm}

\caption{(Color online) Numerical time-resolved KER
and electron spectra of channel 
$\text{CO}^{+} \to \text{CO}^{2+} (\;^3 \Sigma^-) +e^-$.
(a) Time-resolved KER spectrum.  It is distributed broadly in energy 
due to the repulsive potential of the final state.    
In this case, the peak intensity increases over time until it is converged,
i.e. the state $\text{CO}^+$ is completely depopulated.
(b) Time-resolved electron spectrum. It is slightly broader than
the corresponding KER spectrum. Both spectra converge
before 100 fs, and no mirror image relation is found.
}
\label{fig:COspectra}
\end{figure}

The numerical results are shown in Fig.~\ref{fig:COspectra}:
panels (a) and (b) show
the time-resolved KER spectrum and the
time-resolved electron spectrum, respectively.
Both time-resolved spectra finish development within
100 fs, and the electron spectrum is always broader than
the KER spectrum.
We already understand that the breakdown of the mirror image relation
is due to the fact that $|\psi_d (t) \rangle$ is a coherent superposition
of different vibrational levels and due to the finite lifetime broadening.
How do the two spectra differ in their time-resolved developments?
In order to answer this question, we introduce the idea of 
``finite difference of spectra," which also allows us to relate
the spectral developments to the motion of $|\psi_d(t)\rangle$.
The finite difference of spectra is defined as
\begin{equation}
\label{eq:fdspectra}
\frac{\Delta \sigma}{\Delta t} = \frac{\sigma(t_2)-\sigma(t_1)}{t_2-t_1}\;.
\end{equation}
We  will set $t_2=t$ and $t_1=t-1\,\text{fs}$, so the finite
difference of the time-resolved spectra is merely the difference of the spectra 
between $t-1\,\text{fs}$ and $t$.

\begin{figure}
\centering 
\includegraphics[width=8cm,height=5.5cm]{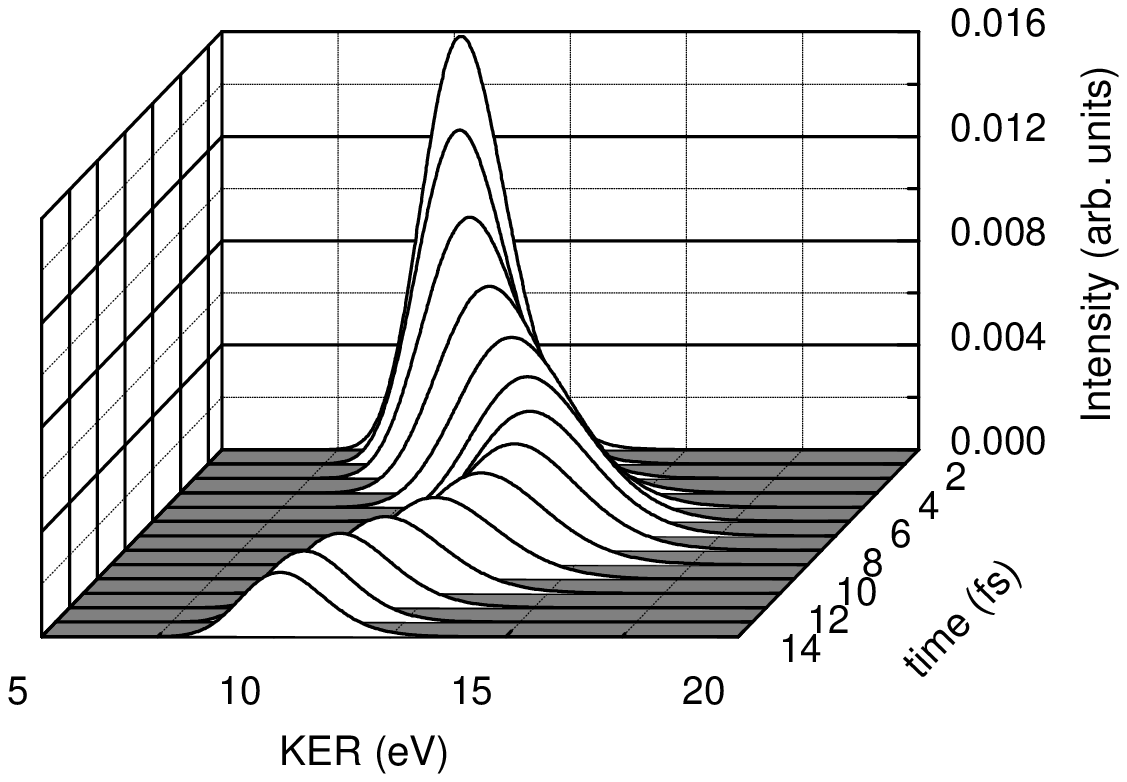} 
\vspace{-0.5cm}
\includegraphics[width=8cm,height=5.5cm]{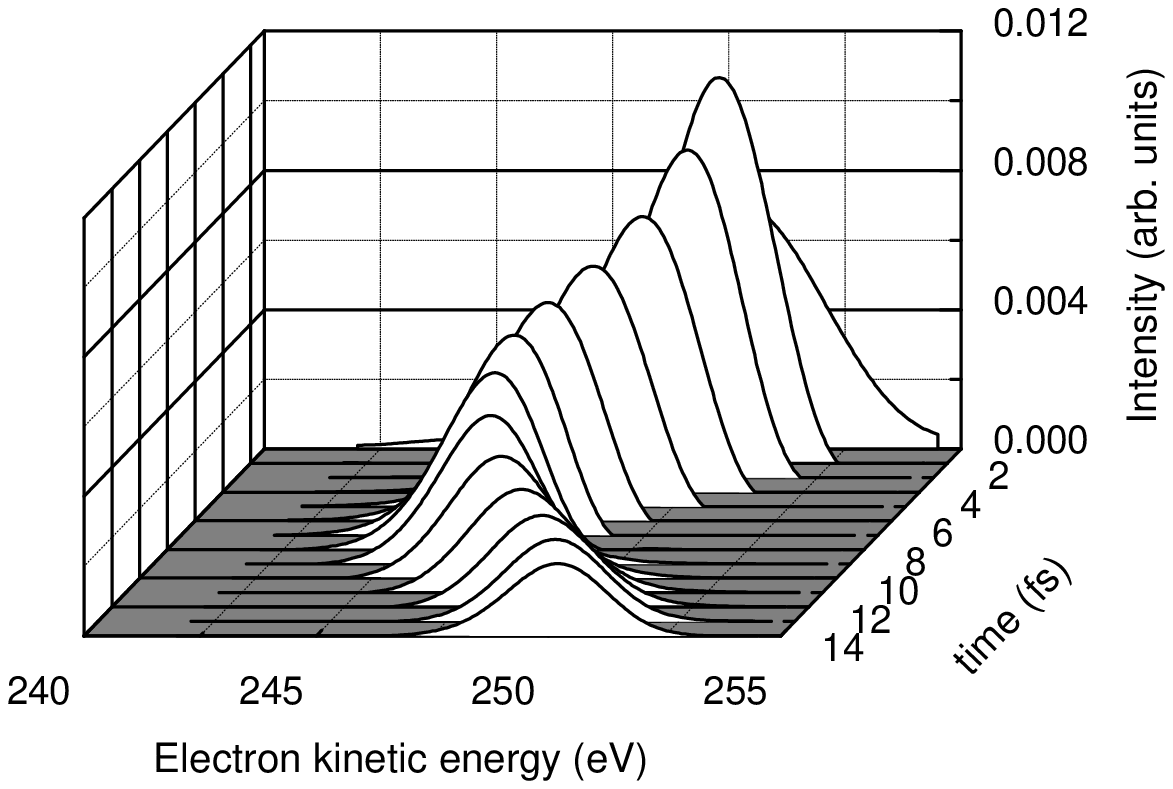} 
\vspace{-0.5cm}

\caption{Finite difference of the time-resolved KER
spectrum, panel (a), and electron spectrum, panel (b), within the first 14 fs. 
The development of the spectrum in panel (a) can be
related to the motion of $|\psi_d(t)\rangle$: 
the period of oscillation, the depopulation of state $d$,
and the dispersion of the wave packet.
Finite difference of time-resolved electron spectrum (b)
shows a strange development within the first 2 fs:
a broad distribution followed by a narrow distribution.
This strange development can be explained by the
interference effect, which comes from the final state wave packet.
}
\label{fig:spectra3D}
\end{figure}

For the time-resolved KER spectrum, the finite differences  
are shown in Fig.~\ref{fig:spectra3D}(a).  
The peak swings back-and-forth in the energy interval
from 10 to 13.4 eV, which correspond to 
the KER spectrum from $|\psi_d(t)\rangle$ 
around the classical turning points.
The oscillation period is 14 fs and is the same as
the period of wave packet motion on the potential $V_d$.
In addition, the depopulation of state $d$ can be observed by
the monotonous decrease of the area of the peak over time. 
The peak also continuously spreads wider over the 
KER energy, which is an evidence for the dispersion of $|\psi_d\rangle$.  
Overall,  measuring the time-resolved KER spectrum allows
one to trace the wave packet's motion on the intermediate state. 

\begin{figure}
\centering 
\includegraphics[width=8cm,height=6cm]{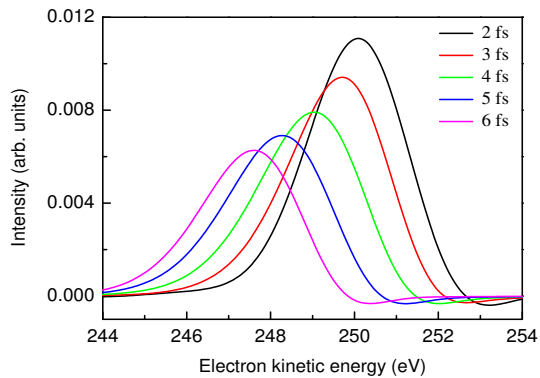} 
\vspace{-0.5cm}

\caption{(Color online) Zoom of Fig.~\ref{fig:spectra3D} (b).
The peak moves to lower electron energy within 2 to 6 fs,
and there is always a negative tail at the right hand side
of the peak. This negative value comes from the interference
term $2 \; \text{Im}[\langle \psi_f(E_e,t) |W_{d \to f}|\psi_d(t)\rangle]$, see text.  }
\label{fig:zoomin}
\end{figure}

The differential electron spectrum
depicted in Fig.~\ref{fig:spectra3D}(b) also 
indicates the oscillation period of $|\psi_d\rangle$.
However, the development of the peak in panel (b) 
is rather nonintuitive.
Take the peak at 1 fs and at 2 fs for example: 
the peak is very broad in the beginning 
but then quickly becomes narrow.  
This curious development is related to the
physical meaning of the time-resolved electron spectrum.
It is a record of the final state population in time,
see Eq.~(\ref{eq:tdespect}).  From Eq.~(\ref{eq:mastereqf}),
we know that $|\psi_f(E_e,t)\rangle$ is generated from
$|\psi_d(t)\rangle$ and is then propagated by $\hat{H}_f+E_e$.
Therefore, there is a unique interference between the 
propagated $|\psi_f(E_e,t)\rangle$ and the incoming
source term $W_{d \to f}|\psi_d (t)\rangle$.  This 
interference sometimes can lead to the final state's depopulation
if the source term has a different phase than the final
state wave packet.
This phenomenon leaves a trace
which becomes more evident if we zoom in to Fig.~\ref{fig:spectra3D}(b),
see Fig.~\ref{fig:zoomin},
When $t$ increases, the peak moves toward smaller electron energy,
and there is a negative tail on the right hand side of the peak.
This negative part is caused by a destructive interference,
appearing only in the electron spectrum but not in the KER spectrum.  
In fact, we can take the finite difference of the spectrum
to the limit $\Delta t \to 0$, i.e. taking the time derivative of the
spectrum. The time-derivative of the KER spectrum gives
$\frac{\partial}{\partial t} \sigma_{\text{KER}} = 2\pi | \langle E_f | W_{d \to f} | \psi_d (t) \rangle|^2$
while the electron spectrum gives
$\frac{\partial}{\partial t} \sigma_{e} =2 \; \text{Im}[\langle \psi_f(E_e,t) |W_{d \to f}|\psi_d(t)\rangle]$.
This shows that partial relative phase information can be extracted from
the time-derivative electron spectrum, but not from the time-derivative
KER spectrum.

\section{Conclusion}

For an autoionization process followed
by fragmentation, we discuss in detail the fully time-dependent approach
for calculating the time-resolved KER spectrum, proposed in \cite{Chiang11},
and compare with the time-resolved electron spectrum.
The physical meaning of the former
is an accumulated generalized Franck-Condon
factor \cite{Chiang11}, while the latter is given 
by the final state population \cite{Elke96}.
The two spectra also develop differently in time
which can be observed from the finite difference
of the spectra.  The finite difference of the KER spectrum
can be related to the motion of
the nuclear wave packet on the intermediate electronic state:
the oscillation period, depopulation of the intermediate
state, and even the dispersion of the wave packet.
On the other hand, the finite difference of the time-resolved
electron spectrum provides the relative phase information
between the wave packets on the final and intermediate states,
$|\psi_f(E_e,t)\rangle$ and $|\psi_d(t)\rangle$, and
exhibits destructive interference
between the source term $W_{d \to f}|\psi_d(t)\rangle$
and the propagated $|\psi_f(E_e,t)\rangle$.  The two time-resolved
spectra together provide complementary information on tracing
the motion of $|\psi_d(t)\rangle$.

Two model examples are employed to
illustrate problems of the mirror image relation
between KER and electron spectra, because the classical
picture underlying this relation
ignores the initial energy distribution of the intermediate state $d$
(or photoelectron) as well as the finite lifetime broadening of
the electron spectrum.  
It is found that the mirror image of the electron spectrum
coincides with the KER spectrum if and 
only if the populated levels of $d$ are quasi-degenerate
and the decay width is rather small.
Such cases have been found in ICD \cite{Jahnke04} or ICD after Auger \cite{Kreidi08_1},
but not for a molecular Auger process \cite{Weber01,Weber03}.
In conclusion, being formally identical to the flux analysis \cite{Beck00,Simona03}
and time-independent approach \cite{Nimrod01} when $t \to \infty$,
the fully time-dependent approach not only provides
remarkably transparent physical insights of the spectra
but also allow us to relate the theory for fragmentation
to the usual pump-probe experiments \cite{Jiang10_2,Bocharova11}.

\section*{Acknowledgements}

Y.C.\,C. thanks Graduate College 850 for financial support.  F.\,O., H.-.D.\,M., and L.S.\,C. thank 
the Deutsche Forschungsgemeinschaft (DFG) for financial support.

\section*{Appendix A: Derivation of Coincidence Spectrum}

The goal of this section is to show that the coincidence spectrum 
obtained from Eq.~(\ref{eq:tdcoincidence}) at $t \to \infty$
is indeed equivalent to the time-independent method reported in Ref.
\cite{Nimrod01}, which is under the condition of broad-band excitation.
Also we would like to prove that Eq.~(\ref{eq:tdker}) is equivalent 
to the flux analysis reported in \cite{Beck00,Simona03} at $t \to \infty$.
To begin with, we need the solution of the time-dependent coefficients
from Eq.~(\ref{eq:exp-psi}). They are
\begin{align}
c_{n_d}(t) &= e^{-i\varepsilon_{n_d}t}\,c_{n_d}(0)\;\theta(t)
\label{eq:sol_cnd}
\\
c_{E_f}(E_e,t) &= \sum_{n_d} \frac{\langle E_f | \hat{W}_{d \to f} | n_d)}{E_f+E_e-\varepsilon_{n_d}}
\, c_{n_d}(0) \left[ e^{-i(E_f+E_e)t} - e^{-i\varepsilon_{n_d}t} \right] \theta(t)
\label{eq:sol_cEf}
\end{align}
where $\theta(t)$ is Heaviside's step function.
For $t<0$, the coefficients vanish because 
only the electronic ground state $i$ is populated
before the ionization process.  Notice that $\varepsilon_{n_d}=E_{n_d}-i \:\Gamma_{n_d}/2$ 
and the decay rate $\Gamma_{n_d}$ is always positive. 

Using Eq.~(\ref{eq:coefficient_f}) and $c_{E_f}(E_e,t)$ from Eq.~(\ref{eq:sol_cEf}),
the final coincidence spectrum reads
\begin{equation}
\sigma(E_\text{KER},E_e) = \lim_{t \to \infty} |c_{E_f}(E_e,t)|^2
= \left|\, \sum_{n_d} \frac{\langle E_f | \hat{W}_{d \to f} | n_d)\, c_{n_d}(0)}%
{E_\text{KER}+V_f^\infty+E_e-\varepsilon_{n_d}} \,\right|^2 \;.
\label{eq:coincidence1}
\end{equation}
This expression is identical with the time-independent 
approach reported in \cite{Nimrod01}.
The next step is to prove that Eq.~(\ref{eq:coincidence1}) can be obtained via
the flux analysis \cite{Beck00}.
We follow here the arguments given in 
Sec.~8.6.3 of \cite{Beck00} with the expression
\begin{equation}
\sigma(E_\text{tot},E_e) = \frac{1}{2\pi} \int_{-\infty}^{+\infty}\!\!\text{d}t'\:e^{-iE_\text{tot}t'}
\int_{-\infty}^{+\infty}\!\!\text{d}t''\:e^{iE_\text{tot}t''}\:
\langle\Psi_\text{tot}(t')|\hat{F}|\Psi_\text{tot}(t'')\rangle
\label{eq:smat-from-flux}
\end{equation}
where $E_{\text{tot}}$ is the total energy of the system and
$\hat{F}$ is a so-called \emph{flux operator}.  It measures, by definition,
the flux into a certain channel, here into the final electronic state $f$
augmented with an emitted electron of energy $E_e$.  
The amount of wave function density which has already reached
the channel is given by the expectation value of the 
projection operator
\begin{equation}
\hat{\Theta} = |\phi_f,E_e \rangle\langle\phi_f,E_e| \;,
\end{equation}
and the time derivative of the expectation value is used to define our flux operator
\begin{equation}
\langle\Psi_\text{tot}(t)|\hat{F}|\Psi_\text{tot}(t)\rangle
=\frac{\text{d}}{\text{d}t}\,\langle\Psi_\text{tot}(t)|\hat{\Theta}|\Psi_\text{tot}(t)\rangle
= \langle\Psi_\text{tot}(t)|i(\hat{H}_\text{tot}^\dagger \hat{\Theta} -
\hat{\Theta}\hat{H}_\text{tot})|\Psi_\text{tot}(t)\rangle \:,
\end{equation}
so $\hat{F} = i(\hat{H}^\dagger_\text{tot} \hat{\Theta} - \hat{\Theta} \hat{H}_\text{tot})$.
Our total Hamiltonian including the electronic components has the form
\begin{align}
\hat{H}_\text{tot} &= (\hat{H}_d - \tfrac{i}{2}\hat{\Gamma}_d) \otimes |\phi_d\rangle\langle\phi_d|
 + \int\!\text{d}E_e\;\hat{W}_{d \to f} \otimes |\phi_f,E_e\rangle\langle\phi_d|
\nonumber\\
&+ \int\!\text{d}E_e\;(\hat{H}_f+E_e) \otimes |\phi_f,E_e\rangle\langle\phi_f,E_e|
\:.
\label{eq:htot}
\end{align}
If we insert this into the definition of the flux operator,
we quickly arrive at
\begin{equation}
\hat{F} = i\bigl(\hat{W}^\dagger_{d \to f}\otimes |\phi_d\rangle\langle\phi_f,E_e| 
- \hat{W}_{d \to f}\otimes|\phi_f,E_e\rangle\langle\phi_d| \bigr)
= \hat{F}_1 + \hat{F}_1^\dagger
\end{equation}
with $\hat{F}_1=i\hat{W}^\dagger_{d \to f}\otimes|\phi_d\rangle\langle\phi_f,E_e|$.

Rewriting Eq.~(\ref{eq:smat-from-flux}) with $\hat{F}_1$ and 
using the expansions from Eq.~(\ref{eq:exp-psi}),
we arrive at
\begin{align}
\sigma(E_\text{tot},E_e) 
&= \frac{i}{2\pi} \int_{-\infty}^{+\infty}\!\!\text{d}t'\:e^{-iE_\text{tot}t'}
\int_{-\infty}^{+\infty}\!\!\text{d}t''\:e^{iE_\text{tot}t''}\:
\langle\psi_d(t')| \hat{W}^\dagger_{d \to f}|\psi_f(t'')\rangle + \text{c.c.}
\nonumber \\
&= \frac{i}{2\pi} \sum_{n_d} \int\!\text{d}E_f\: (n_d|\hat{W}^\dagger_{d\to f}|E_f\rangle
\!\!\int\limits_{-\infty}^{+\infty}\!\!\text{d}t'\:e^{-i E_\text{tot}t'} c_{n_d}(t')^*
\!\!\int\limits_{-\infty}^{+\infty}\!\!\text{d}t''\:e^{i E_\text{tot}t''} c_{E_f}(E_e,t'')
\,+\,\text{c.c.} 
\label{eq:smat-from-ft}
\end{align}
This equation requires the Fourier transforms of 
$c_{n_d}(t)$ and $c_{E_f}(E_e,t)$. Using Eqs.~(\ref{eq:sol_cnd}, \ref{eq:sol_cEf}), the former is
\begin{equation}
\int_{-\infty}^{+\infty}\!\!\text{d}t\:e^{i E_\text{tot}t} c_{n_d}(t)
= \int_{-\infty}^{+\infty}\!\!\text{d}t\:e^{i (E_\text{tot}-\varepsilon_{n_d})t} c_{n_d}(0) \theta(t)
= \frac{i\,c_{n_d}(0)}{E_\text{tot}-\varepsilon_{n_d}} \;,
\end{equation}
and the latter yields
\begin{align}
&\int_{-\infty}^{+\infty}\!\!\text{d}t\:e^{i E_\text{tot}t} c_{E_f}(E_e,t)
\nonumber\\
&\qquad= \sum_{n_d} \frac{\langle E_f | \hat{W}_{d \to f} | n_d)}{E_f+E_e-\varepsilon_{n_d}} \, c_{n_d}(0)
  \left[ \frac{i}{E_\text{tot}\!-\!E_f\!-\!E_e} + \pi\,\delta(E_\text{tot}\!-\!E_f\!-\!E_e)
  - \frac{i}{E_\text{tot}\!-\!\varepsilon_{n_d}} \right]
\nonumber\\
&\qquad= \sum_{n_d} \langle E_f | \hat{W}_{d \to f} | n_d) \, c_{n_d}(0)
\left[ \frac{i}{(E_\text{tot}\!-\!E_f\!-\!E_e)(E_\text{tot}\!-\!\varepsilon_{n_d})}
+\frac{\pi\,\delta(E_\text{tot}\!-\!E_f\!-\!E_e)}{E_\text{tot}\!-\!\varepsilon_{n_d}}
\right]
\nonumber\\
&\qquad= \left[ \frac{i}{E_\text{tot}\!-\!E_f\!-\!E_e} + \pi\,\delta(E_\text{tot}\!-\!E_f\!-\!E_e) \right]
\sum_{n_d} \frac{\langle E_f | \hat{W}_{d \to f} | n_d)}{E_\text{tot}\!-\!\varepsilon_{n_d}}\, c_{n_d}(0) \;.
\end{align}
We can now rearrange Eq.~(\ref{eq:smat-from-ft}) to obtain
\begin{align}
\sigma(E_\text{tot},E_e) &= \frac{i}{2\pi} \int\!\text{d}E_f\:
\sum_{n_d}\,(n_d|\hat{W}^\dagger_{d\to f}|E_f\rangle\,\frac{-i\,c_{n_d}(0)^*}{E_\text{tot}-\varepsilon_{n_d}^*}
\nonumber\\
&\quad\times
\left[ \frac{i}{E_\text{tot}\!-\!E_f\!-\!E_e} + \pi\,\delta(E_\text{tot}\!-\!E_f\!-\!E_e) \right]
\;
\sum_{n_d'}\,\frac{\langle E_f | \hat{W}_{d \to f} | n_d)}{E_\text{tot}\!-\!\varepsilon_{n_d}}\, c_{n_d}(0)
\;+\;\text{c.c.}
\nonumber\\
&= \frac{1}{2\pi} \int\!\text{d}E_f\: \left[ \frac{i}{\cdots} + \pi\,\delta(\cdots) \right] \left|
\sum_{n_d} \frac{\langle E_f | \hat{W}_{d \to f} | n_d)}{E_\text{tot}\!-\!\varepsilon_{n_d}}\, c_{n_d}(0)
\right|^2 \;+\;\text{c.c.}
\nonumber\\
&= \int\!\text{d}E_f\: \delta(E_\text{tot}\!-\!E_f\!-\!E_e) \left|
\sum_{n_d} \frac{\langle E_f | \hat{W}_{d \to f} | n_d)}{E_\text{tot}\!-\!\varepsilon_{n_d}}\, c_{n_d}(0)
\right|^2
\nonumber\\
&= \left| \sum_{n_d} \frac{\langle E_f|\,
\hat{W}_{d \to f}\,|\,n_d) c_{n_d}(0)}{E_\text{KER}+V_f^\infty+E_e-\varepsilon_{n_d}} 
\right|^2
\end{align}
with $E_{\text{tot}}=E_e+E_{\text{KER}}+V_f(\infty)$. 
Thus the flux analysis \cite{Beck00} indeed can be applied to simulate the coincidence spectrum.
 
The remaining piece is to demonstrate that the coincidence spectrum can
be further modified as in Ref.~\cite{Simona03} when a
complex absorbing potential (CAP) \cite{Leforestier83,Kosloff86,Neuhauser89} 
is applied for improving numerical efficiency.
Since only the final electronic state is repulsive, the CAP
is only needed for state $f$.  Thus, the total Hamiltonian
becomes
\begin{equation}
\hat{H}'_\text{tot} = \hat{H}_\text{tot} - i\hat{W}_{\text{CAP}}
\end{equation}
with
\begin{equation}
\hat{W}_{\text{CAP}} = \hat{W}_R \otimes \int \text{d}E_e |\phi_f, E_e \rangle \langle \phi_f, E_e | \:.
\end{equation}
Following \cite{Beck00}, if the CAP is sufficiently weak (so that it causes
only negligible reflections), the wave function outside the CAP region won't be changed
if the propagation is done by $\hat{H}'_\text{tot}$ instead of $\hat{H}_\text{tot}$.
The only thing needing modification is the flux operator, which now reads
\begin{align}
\hat{F}
&= i\left( (\hat{H}_\text{tot}^{\prime\dagger}-i\hat{W}_{\text{CAP}})\hat{\Theta}
         - \hat{\Theta}(\hat{H}'_\text{tot}+i\hat{W}_{\text{CAP}}) \right)
\nonumber\\
&= 2\hat{W}_{\text{CAP}}\hat{\Theta} 
 + i(\hat{H}_\text{tot}^{\prime\dagger} \Theta - \Theta \hat{H}'_\text{tot}) \;.
\end{align}
The result is that Eq.~(\ref{eq:smat-from-flux}) is split into two parts:
\begin{align}
\sigma(E_\text{tot})
&= \frac{1}{\pi} \int_{-\infty}^{+\infty}\!\!\text{d}t' \int_{-\infty}^{+\infty}\!\!\text{d}t''\:
   e^{-i E_\text{tot}(t'-t'')}\,\langle\Psi_\text{tot}(t')|\hat{W}_{\text{CAP}}\hat{\Theta}|\Psi_\text{tot}(t'')\rangle
\nonumber\\
&+ \frac{1}{2\pi} \int_{-\infty}^{+\infty}\!\!\text{d}t' \int_{-\infty}^{+\infty}\!\!\text{d}t''\:
   e^{-i E_\text{tot}(t'-t'')} \left( \frac{\text{d}}{\text{d}t'} + \frac{\text{d}}{\text{d}t''} \right)
   \langle\, \Psi_\text{tot}(t') \,|\, \hat{\Theta} \,|\,
             \Psi_\text{tot}(t'') \,\rangle
\end{align}
The second part depends on the behavior of
$\langle \Psi_\text{tot}(t') | \hat{\Theta} | \Psi_\text{tot}(t'')\rangle
= \langle \psi_f(E_e,t') | \psi_f(E_e,t'') \rangle$
for $t',t'' \to \pm\infty$; but it vanishes
due to the initial conditions,
$\psi_f(E_e,t)=0$ for $t<0$, and because of the CAP, as $\psi_f(E_e,t)\to 0$ for $t\to\infty$.
Here we have assumed that the final state potential is 
purely repulsive. Hence all parts of the wave function will reach
the CAP, and there is no bound part of the wave function.
(For non-vanishing $ |\psi_f(E_e,t)\rangle$ at $t \to \infty$,
the correction can be found in Ref.~\cite{Sukiasyan02}.) 
This leaves only the first part; noting that
$\langle \Psi_\text{tot}(t') | \hat{W}_\text{CAP} \hat{\Theta} | \Psi_\text{tot}(t'')\rangle
= \langle \psi_f(E_e,t') | \hat{W}_R | \psi_f(E_e,t'') \rangle$, it's clear
that the lower integral boundaries can be replaced by zero, leaving
\begin{align}
\sigma(E_\text{tot}) 
&= \frac{1}{\pi} \int_0^{\infty}\!\!\text{d}t' \int_0^{\infty}\!\!\text{d}t''\:
e^{-i E_\text{tot}(t'-t'')}\,\langle \psi_f(E_e,t') | \hat{W}_R | \psi_f(E_e,t'') \rangle
\nonumber \\
&= \frac{2}{\pi} \text{Re}\int_0^{\infty}\!\!\text{d} \tau  \: e^{i E_\text{tot}\tau}
\int_0^{\infty}\!\!\text{d}t \: \langle \psi_f(E_e,t) | \hat{W}_R | \psi_f(E_e,t+\tau) \rangle 
\:,
\end{align}
where $t=t'$ and $\tau=t''-t'$.  This expression is identical with Ref. \cite{Simona03}.

\section*{Appendix B: Discrete-function of the continuum}

The computation of the KER spectrum via Eq.~\ref{eq:tdker} requires a representation
of the $\delta$-normalized continuum state $|E_f\rangle$, or more precisely, of the
projector $|E_f\rangle\langle E_f|$. This projector can be expressed by
a $\delta$-function and, in turn, by a representation thereof
\begin{equation}
|E_f\rangle \langle E_f| = \delta (E_f-H_f) = \frac{1}{\pi} \lim_{\epsilon \to 0^+} \text{Im} 
\frac{-1}{E_f-H_f+i\epsilon}
\end{equation}
The constant $i\epsilon$ can be replaced \cite{Seideman92} by a complex absorbing potential (CAP)
\cite{Riss93}. To this end the CAP-augmented Hamiltonian is introduced
\begin{equation}
\tilde{H_f}=H_f-i\eta W_{\text{CAP}} \;,
\end{equation}
where $W_{\text{CAP}}$ is a non-negative potential-like function which
vanishes in the interior region but increases for large distances. 
We have used
\begin{equation}
W_{\text{CAP}}(R)=(R-R_c)^3 \; \Theta(R-R_c) \;,
\end{equation}
where $\Theta$ denotes the Heaviside step function and $R_c$ is the
point where the CAP is switched on. The CAP strength $\eta$ must be chosen
strong enough such that also high energy wave functions are absorbed
before they reach the end of the grid, but small enough to ensure that
the CAP does not introduce significant reflections \cite{Riss96}.
Note that the spectrum of the CAP-augmented Hamiltonian $\tilde{H_f}$
is purely discrete
\begin{equation}
\tilde{H_f}\phi_j = \tilde{E}^f_j \phi_j
\end{equation}
where the eigenfunctions $\phi_j$ are normalized with respect
to the symmetric scalar product $(\phi_j|\phi_k)=\delta_{jk}$. 
The complex eigenvalues $\tilde{E}^f_j$ have negative imaginary parts.
Using Eq.~\ref{eq:tdker}, we may now write the KER spectrum as
\begin{equation}
\sigma_{\text{KER}}(E_{\text{KER}},t) 
= 2 \pi \int_0^t \text{d} t' \; \langle \psi_d (t') | \hat{W}^\dagger_{d \to f}  \; \frac{1}{\pi} \; \text{Im} \frac{-1}{E_f-\tilde{H}_f} \; \hat{W}_{d \to f} | \psi_d (t') \rangle 
\end{equation}
where  $E_{\text{KER}}=E_f-V_f^{\infty}$.
Performing the inverse by diagonalisation yields
\begin{equation}
\sigma_{\text{KER}}(E_{\text{KER}},t)
= -2 \sum_j \; \text{Im} \int_0^t \text{d} t' \; \langle \psi_d (t') | \hat{W}^\dagger_{d \to f} | \phi_j ) \; \frac{1}{E_f-\tilde{E}^f_j} \;
( \phi_j | \hat{W}_{d \to f} | \psi_d(t') \rangle \;.
\end{equation}
This equation is our working equation. It is very stable and allows us to vary $\eta$
over more than three orders of magnitude without virtually changing the computed KER spectrum.
Only when $\eta$ is too small we observed the artificial higher frequency oscillation of high energy part
of the KER spectrum and when $\eta$ is too large 
there appear lower-frequency oscillations of the low energy part of the KER spectrum.

\bibliographystyle{phaip}
\bibliography{ker}

\begin{thebibliography}{10}

\bibitem{Nasrin08}
N.~{Mirsaleh-Kohan}, W.~D. {Robertson}, and R.~N. {Compton},
\newblock Mass Spectrometry Reviews {\bf 27}, 237 (2008).

\bibitem{Jagutzki02}
O.~Jagutzki et~al.,
\newblock IEEE Transactions on Nuclear Science {\bf 49}, 2477 (2002).

\bibitem{Schinke}
R.~{Schinke},
\newblock {\em Photodissociation Dynamics},
\newblock Press Syndicate of the University of Cambridge, Cambridge, 1993.

\bibitem{Purnell94}
J.~Purnell, E.~M. Snyder, S.~Wei, and A.~W.~C. Jr.,
\newblock Chem. Phys. Lett. {\bf 229}, 333 (1994).

\bibitem{Stapelfeldt95}
H.~Stapelfeldt, E.~Constant, and P.~B. Corkum,
\newblock Phys. Rev. Lett. {\bf 74}, 3780 (1995).

\bibitem{Lezius98}
M.~Lezius, S.~Dobosz, D.~Normand, and M.~Schmidt,
\newblock Phys. Rev. Lett. {\bf 80}, 261 (1998).

\bibitem{Eberhardt87}
W.~Eberhardt, E.~W. Plummer, I.~W. Lyo, R.~Carr, and W.~K. Ford,
\newblock Phys. Rev. Lett. {\bf 58}, 207 (1987).

\bibitem{Weber03}
T.~Weber et~al.,
\newblock Phys. Rev. Lett. {\bf 90}, 153003 (2003).

\bibitem{Lenz97}
L.~S. Cederbaum, J.~Zobeley, and F.~Tarantelli,
\newblock Phys. Rev. Lett. {\bf 79}, 4778 (1997).

\bibitem{Robin00}
R.~Santra, J.~Zobeley, L.~S. Cederbaum, and N.~Moiseyev,
\newblock Phys. Rev. Lett. {\bf 85}, 4490 (2000).

\bibitem{Marburger03}
S.~Marburger, O.~Kugeler, U.~Hergenhahn, and T.~M\"oller,
\newblock Phys. Rev. Lett. {\bf 90}, 203401 (2003).

\bibitem{Jahnke04}
T.~Jahnke et~al.,
\newblock Phys. Rev. Lett. {\bf 93}, 163401 (2004).

\bibitem{Nico10_1}
N.~Sisourat et~al.,
\newblock Nature Phys. {\bf 6}, 508 (2010).

\bibitem{Weber01}
T.~Weber et~al.,
\newblock Journal of Physics B: Atomic, Molecular and Optical Physics {\bf 34},
  3669 (2001).

\bibitem{Morin86}
P.~Morin and I.~Nenner,
\newblock Phys. Rev. Lett. {\bf 56}, 1913 (1986).

\bibitem{Menzel96}
A.~Menzel, B.~Langer, J.~Viefhaus, S.~B. Whitfield, and U.~Becker,
\newblock Chem. Phys. Lett. {\bf 258}, 265 (1996).

\bibitem{Elke98}
E.~Pahl, L.~S. Cederbaum, H.-D. Meyer, and F.~Tarantelli,
\newblock Phys. Rev. Lett. {\bf 80}, 1865 (1998).

\bibitem{Faris99}
F.~Gel'mukhanov and H.~\"Agren,
\newblock Phys. Rep. {\bf 312}, 87 (1999).

\bibitem{Simona04}
S.~Scheit et~al.,
\newblock J. Chem. Phys. {\bf 121}, 8393 (2004).

\bibitem{Chiang11}
Y.-C. Chiang, F.~Otto, H.-D. Meyer, and L.~S. Cederbaum,
\newblock Phys. Rev. Lett. {\bf 107}, 173001 (2011).

\bibitem{Jiang10_2}
Y.~H. Jiang et~al.,
\newblock Phys. Rev. A {\bf 81}, 051402(R) (2010).

\bibitem{Bocharova11}
I.~A. Bocharova et~al.,
\newblock Phys. Rev. A {\bf 83}, 013417 (2011).

\bibitem{Elke96}
E.~Pahl, H.-D. Meyer, and L.~S. Cederbaum,
\newblock Z. Phys. D {\bf 38}, 215 (1996).

\bibitem{Eland04}
J.~H.~D. Eland et~al.,
\newblock Journal of Physics B: Atomic, Molecular and Optical Physics {\bf 37},
  3197 (2004).

\bibitem{Carroll02}
T.~X. Carroll et~al.,
\newblock J. Chem. Phys. {\bf 116}, 10221 (2002).

\bibitem{Matsumoto06}
M.~Matsumoto et~al.,
\newblock Chem. Phys. Lett. {\bf 417}, 89 (2006).

\bibitem{Nico10_2}
N.~Sisourat, N.~V. Kryzhevoi, P.~Koloren\v{c}, S.~Scheit, and L.~S. Cederbaum,
\newblock Phys. Rev. A {\bf 82}, 053401 (2010).

\bibitem{Beck00}
M.~H. Beck, A.~J\"ackle, G.~A. Worth, and H.-D. Meyer,
\newblock Phys. Rep. {\bf 324}, 1 (2000).

\bibitem{Simona03}
S.~Scheit, L.~S. Cederbaum, and H.-D. Meyer,
\newblock J. Chem. Phys. {\bf 118}, 2092 (2003).

\bibitem{Kreidi08_1}
K.~Kreidi et~al.,
\newblock Phys. Rev. A {\bf 78}, 043422 (2008).

\bibitem{Nimrod01}
N.~Moiseyev, J.~Zobeley, R.~Santra, and L.~S. Cederbaum,
\newblock J. Chem. Phys. {\bf 114}, 7351 (2001).

\bibitem{Leforestier83}
C.~Leforestier and R.~E. Wyatt,
\newblock J. Chem. Phys. {\bf 78}, 2334 (1983).

\bibitem{Kosloff86}
R.~Kosloff and D.~Kosloff,
\newblock J. Comp. Phys. {\bf 63}, 363 (1986).

\bibitem{Neuhauser89}
D.~Neuhauser and M.~Baer,
\newblock J. Chem. Phys. {\bf 90}, 4351 (1989).

\bibitem{Sukiasyan02}
S.~Sukiasyan and H.-D. Meyer,
\newblock J. Chem. Phys. {\bf 116}, 10641 (2002).

\bibitem{Seideman92}
T.~Seideman and W.~H. Miller,
\newblock J. Chem. Phys. {\bf 96}, 4412 (1992).

\bibitem{Riss93}
U.~V. Riss and H.-D. Meyer,
\newblock J. Phys. B: At. Mol. Opt. Phys. {\bf 26}, 4503 (1993).

\bibitem{Riss96}
U.~V. Riss and H.-D. Meyer,
\newblock J. Chem. Phys. {\bf 105}, 1409 (1996).

\end{thebibliography}

\end{document}